\documentclass[a4paper,twocolumn,11pt]{quantumarticle}
\pdfoutput=1
\usepackage[utf8]{inputenc}
\usepackage[english]{babel}
\usepackage[T1]{fontenc}

\usepackage{tikz}
\usepackage{lipsum}

\usepackage{amsmath,amssymb,bm,braket}
\usepackage{amsthm}
\usepackage{microtype}
\usepackage{hyphenat}

\newcommand{\Hcal}{\mathcal{H}}
\newcommand{\Bcal}{\mathcal{B}}
\newcommand{\Tr}{\mathrm{Tr}}
\newcommand{\id}{\mathbb{I}}
\newcommand{\erf}{\operatorname{erf}}

\newtheorem{lemma}{Lemma}
\usepackage{etoolbox}
\AtEndEnvironment{lemma}{\vspace{-0.8\baselineskip}}

\usepackage{caption}
\usepackage[colorlinks=true,allcolors=blue]{hyperref}

\begin{document}

\title{Operationally induced preferred basis in unitary quantum mechanics}
\author{Vitaly Pronskikh}
\affiliation{Fermi National Accelerator Laboratory, Batavia, USA email: \texttt{vspron@fnal.gov}}

\begin{abstract}
\begin{abstract}
The preferred-basis problem and the definite-outcome aspect of the measurement problem persist even when the detector is modeled unitarily. Experimental data are represented in a Boolean event algebra of mutually exclusive records, while the theoretical description employs a noncommutative operator algebra with continuous unitary symmetry. This change of mathematical structure constitutes the core of the ``cut'': a necessary interface from group-based kinematics to set-based counting.

In the Operationally Induced Preferred Basis (OIPB) framework, the basis relevant for recorded outcomes is not fixed by the system Hamiltonian but induced by the measurement interface---the detector channel together with its coarse-grained readout. The Born rule follows from Gleason-type uniqueness (Gleason for projections in $d>2$ and Busch's extension for POVMs including $d=2$), as the unique probability measure consistent with additivity over exclusive events and basis-independence of the unitary sector. A compact qubit--pointer model yields an induced unsharp POVM $E_{\pm}=\frac12(\id\pm\eta\sigma_z)$ with sharpness $\eta$ fixed by pointer resolution, explicitly demonstrating detector-induced basis selection.

OIPB aligns with decoherence and operational theories while diverging from collapse models (no spontaneous reductions) and the Many-Worlds Interpretation (no ontological branching). Empirical distinctions arise through POVM tomography, Wigner-friend incompatibility tests, and superposition stability bounds. Nested-observer paradoxes are resolved by a non-composability lemma: joint assignment of outcome propositions is possible only if a joint instrument exists. This relocates the origin of randomness to the stochasticity of the interface rules.
\end{abstract}
\end{abstract}

\maketitle 

\section{Introduction}
The measurement problem is traditionally characterized by the tension between the deterministic, unitary evolution of the state vector (von Neumann's Process 2) and the stochastic, non-unitary reduction associated with observation (Process 1)~\cite{vonNeumann1932}. While the program of Environment-Induced Decoherence has successfully explained the dynamical suppression of interference terms—effectively transitioning from a pure state to an improper mixture—it formally fails to derive the selection of a unique outcome without assuming a pre-existing partition of the Hilbert space. This 'preferred basis problem' indicates that unitary dynamics alone are insufficient to account for the definitive 'and-or' transition observed in experiments. Consequently, a complete description requires a rigorous formalization of the boundary conditions that constrain the readout protocol.

This paper develops a methodological/foundational proposal designed to remain within standard quantum measurement theory. The central claim is that the preferred basis relevant for recorded outcomes is not extracted from the system Hamiltonian in isolation; it is induced by the readout scheme (coupling plus readout coarse-graining). A second claim addresses the usual ``why the square?'' objection: the Born trace rule is not treated as an arbitrary protocol, but as the unique interface measure compatible with additivity over exclusive outcomes and basis-independence of the unitary kinematics, as formalized by Gleason-type theorems.

Fock advocated a clear operational approach to quantum experiments, criticizing Bohr’s complementarity for conflating the description of quantum systems with the instruments used to observe them~\cite{fock1957,fock1958}. A key element of his framework is the division of any experiment into three distinct stages: preparation of the initial state, unitary evolution (the working stage), and registration of the outcome. 

This tripartite structure provides the natural foundation for the Operationally Induced Preferred Basis (OIPB). In OIPB, the preferred basis is not dictated by the system Hamiltonian alone, but emerges from the interplay between preparation, unitary dynamics, and the specific registration interface implemented by the detector.

As Fock articulated, ``If a subdivision of the experimental arrangement in a preparatory part, a working part and a registration part is possible, one can vary the last stage of experiment and obtain probability distributions referring to the same initial state,'' thereby demonstrating the wave function's independence from measurement specifics and enabling abstraction to a state concept invariant under registration variations~\cite{fock1957}. 

This division anticipates OIPB's core mechanism, wherein the preferred basis emerges not from intrinsic dynamical privileges but from the interplay of operational constraints across these stages---preparation selects initial superpositions, evolution propagates them unitarily, and registration enforces classical-like decoherence, collectively inducing a basis aligned with measurement outcomes. By grounding quantum states in such empirically delineated processes, Fock's schema resolves interpretive tensions akin to the preferred basis problem, paving the way for OIPB's resolution of Born rule derivations and non-collapse dynamics within a fully operational paradigm. 

While David Wallace's decision-theoretic derivation of the Born rule in the Everettian framework~\cite{wallace2012} shares much with OIPB and Fock's operational schema---particularly in using decoherence to select a preferred basis through environmental interactions in the unitary stage, thereby linking probabilities to the preparation-propagation-registration interplay without collapse---it parts ways in its multiverse ontology, which OIPB avoids in favor of empirical invariance over branching worlds. Likewise, Dennis Dieks' decoherence work~\cite{dieks1989} echoes Fock's tripartite structure by showing how entanglements at the registration stage dampen off-diagonal density matrix terms, producing robust pointer states that address measurement issues via operational constraints rather than inherent dynamics; yet Dieks remains ontologically flexible, even accommodating partial collapses, whereas OIPB, true to Fock's realism, commits to a collapse-free view that derives states purely from delineated experiments, sidestepping any lingering subjectivity in outcomes.

Several established frameworks articulate a closely related theoretical--operational division in explicitly physicist terms. In the operational approach of Davies and Lewis, a measurement is formalized as an instrument, i.e., a completely positive map--valued measure that outputs both a classical outcome distribution and a conditional state update, thereby making the quantum-to-classical outcome map part of the theory rather than an informal appendage \cite{DaviesLewis1970}. In operator-algebraic measurement models, the emergence of records is encoded by distinguished commuting macroscopic observables (or ``phase cells'') that effectively carry a Boolean event structure for pointer positions \cite{Hepp1972,Sewell2005}.

In a complementary direction, Landsman’s Bohrification program systematically organizes classical contexts as commutative $C^*$-subalgebras inside a noncommutative algebra of observables, thereby formalizing the idea that empirical content is accessed through commutative (classical) descriptions \cite{Landsman2016Bohrification}. The present approach is compatible with these lines of work but shifts emphasis to the induced instrument/POVM as the physically specified mapping that selects the operational event algebra (hence the preferred basis) while retaining standard unitary dynamics in the theoretical sector.  In contrast, we frame the induced instrument/POVM as the physical boundary that actively selects the operational event algebra, thereby deriving the preferred basis from contingent detector properties rather than from a priori classical contexts. Both frameworks reject the idea that the quantum-classical divide can be ignored or unitarily eliminated, but where~\cite{Landsman2016Bohrification} locates classicality in the algebraic hierarchy of observables, we locate it in the empirically specified measurement protocol that induces the event algebra relevant for recorded data.

Another related decomposition of complex measurements was advocated in the theoretical--operational analysis of experimental practice, where the phenomenon is accessed only through explicitly modeled preparation and measurement operations under a central theoretical description \cite{Pronskikh2020PhysUsp}. In that scheme, measurement is not a monolithic act but a structured composition of indication (record formation) and data analysis, and the effective reference used for attributing values may be partially constituted at the recognition/interpretation stage. In the present language this motivates treating the instrument (and its induced POVM) as the physically specified bridge: it fixes the admissible outcome algebra and thereby the operationally preferred basis, while the unitary model governs only the pre-interface evolution.

Our approach can be considered ``structural--operational'' (or an algebraic interface protocol) as shorthand for this correspondence-based reading of standard instrument theory. It is realist about (i) the unitary kinematical structure and its symmetries, and (ii) the operational event structure implemented by measurement readout, without committing to a specific microscopic ``object'' ontology beyond the standard formalism.

\section{Kinematic Sectors and Algebraic Constraints}

We define the readout scheme through the contrast between two distinct algebraic structures. Let the theoretical state of the system be described by the non-commutative $C^*$-algebra of bounded operators $\mathfrak{A} = \mathcal{B}(\mathcal{H})$ acting on a Hilbert space $\mathcal{H}$. This sector is governed by the continuous unitary group $SU(N)$, preserving the superposition principle. Conversely, the operational readout of an instrument is constrained to a commutative Boolean algebra $\mathfrak{B}$, representing the set of mutually exclusive, classical events (clicks or counts).

Noncommutativity implies that not all observables admit a single joint classical sample space; joint outcome propositions exist only when the corresponding measurements are jointly measurable. In this sense, the Boolean event algebra realized in an experiment is fixed by the chosen instrument (context), rather than determined by unitary dynamics alone \cite{Heinosaari2016Invited}. The ``Heisenberg Cut'' is hereby formalized not as an arbitrary shift in the observer's placement, but as the singular mapping $\mathcal{I}: \mathfrak{A} \to \mathfrak{B}$. The statistical mapping is represented operationally by an instrument (or POVM together with an update rule), i.e., a CP map--valued measure that outputs a classical outcome and a conditional post-measurement state. This map is not a unitary on the system algebra because it is not invertible at the level of recorded events.

\subsection{Theoretical sector: noncommutative structure and unitary symmetry}

Let $\Hcal$ be the system Hilbert space. Closed-system dynamics is described by a unitary group representation,
\begin{equation}
\rho(t)=U(t)\rho(0)U^\dagger(t), \qquad U(t)=e^{-iHt/\hbar}.
\label{eq:unitary}
\end{equation}
Sharp quantum propositions correspond to projections $P\in\mathcal{P}(\Hcal)$, whose lattice is orthomodular and generally non-distributive \cite{BirkhoffVN1936}. The non-Boolean character is not a philosophical gloss; it is the algebraic expression of superposition and noncommutativity.

\subsection{Operational sector: Boolean event algebra and counting}

Experimental outcomes are represented as elements of a classical event space $(\Omega,\Sigma)$ where $\Sigma$ is a Boolean $\sigma$-algebra. Each run produces one stored label $i\in\Omega$. This is the domain in which frequencies, likelihoods, and confidence intervals are defined. The operational description is therefore set-theoretic and distributive by construction: one does not store ``superpositions of records.''

\subsection{Algebraic incommensurability and the meaning of the cut}

Modeling the detector unitarily enlarges the theoretical sector but does not remove the structural mismatch: to talk about recorded outcomes one must apply the interface map $\mathcal{I}$ to carry the noncommutative description into a Boolean event algebra. The ``cut'' in this paper is precisely that switch of type: from unitary symmetry acting on operator algebras to set-based exclusive events suitable for counting. This is not introduced as an approximation; it is a requirement for formulating empirical claims. Indeed, it is a requirement that no unitary extension of the quantum dynamics can eliminate.

Standard quantum mechanics creates a tension between the continuous unitary group $SU(N)$ governing state evolution and the discrete Boolean lattice $\mathfrak{B}$ governing experimental outcomes. We propose that the measurement ``cut'' is formally defined as the non-unitary map $\mathcal{I}$ interfacing these two distinct kinematic sectors:
\begin{equation}
    \mathcal{I}:\mathfrak{A}_Q\to \mathfrak{B}_{\text{Boolean}},
\end{equation}
where $\mathfrak{A}_{Q}$ is the non-commutative $C^*$-algebra of the quantum system and $\mathfrak{B}_{\text{Boolean}}$ is the commutative Boolean algebra of the detector readout. As shown by the no-go theorems on joint measurability for non-commuting observables~\cite{Heinosaari2016Invited}, there exists no group homomorphism that continuously maps the unitary symmetries of $\mathfrak{A}_{Q}$ onto the discrete logic of $\mathfrak{B}_{\text{Boolean}}$. Consequently, the boundary $\mathcal{I}$ functions as a \textit{topological constraint}, necessitating a singular projection (or ``collapse'') to satisfy the boundary conditions of the operational readout.

A POVM is the standard mathematical representation of this demarcation map. It is a countably additive map $E:\Sigma\to\Bcal(\Hcal)$ with $E(A)\ge 0$ and $E(\Omega)=\id$ \cite{Holevo1982,BuschLahtiMittelstaedt1996}. The Born probability assignment is then
\begin{equation}
p(A|\rho)=\Tr\!\big(\rho\,E(A)\big).
\label{eq:born-povm}
\end{equation}
Our algebraic protocol emphasizes that $E$ is part of the physical specification of a measurement interface: it encodes what the apparatus can stably register as classical events.

\section{Born probabilities as a structural interface constraint}
\label{sec:born}

The question ``why $|\psi|^2$ and not another power?'' becomes: which probability measures on quantum event structures are compatible with additivity over exclusive outcomes and with the basis-independence of the unitary kinematics.

In the sharp case, one seeks a measure $\mu:\mathcal{P}(\Hcal)\to[0,1]$ satisfying $\sum_k \mu(P_k)=1$ for every orthogonal resolution of identity ${P_k}$ \cite{Gleason1957}. Gleason’s theorem states that for $\dim\Hcal\ge 3$,
\begin{equation}
\mu(P)=\Tr(\rho P)
\label{eq:gleason}
\end{equation}
for a unique density operator $\rho$ \cite{Gleason1957}. For $P_k=\ket{k}\!\bra{k}$ and $\rho=\ket{\psi}\!\bra{\psi}$
this reduces to $|\braket{k|\psi}|^2$.

A wide range of proposed ``derivations'' of the Born rule exist, typically requiring assumptions beyond unitary dynamics; a critical overview is given by Vaidman \cite{Vaidman2020BornReview}. Our approach belongs to the Gleason-type uniqueness class: once one requires that the mapping assign additive probabilities to exclusive events and remain consistent under unitary changes of theoretical frame, the trace rule is singled out as the unique probability measure on the induced event structure.

Here we interpret those assumptions not as metaphysical postulates or rationality axioms, but as requirements on the quantum-to-classical mapping: a Boolean event algebra must be produced for records, and the probability assignment must be invariant under the unitary ``frame'' on the theoretical side.

For operational measurements described directly by POVMs (effects), Busch proved an analogous uniqueness statement that covers all finite dimensions, including $d=2$ \cite{Busch2003PRL}. Thus, these uniqueness theorems constrain the correspondence: the trace rule is the unique probability measure that respects additivity over exclusive events while remaining compatible with the unitary sector's basis-independence.

While Gleason's theorem derives the Born rule as the unique consistent measure for dimensions $d > 2$, it notably fails for the qubit case ($d=2$). To resolve this within our algebraic framework, we employ Busch's extension of Gleason's theorem to POVMs. 

For a two-level system, the requirement that the probability measure $\mu(E)$ be continuous and additive over effects yields the trace rule uniquely, provided the measurement is unsharp (as realistic detectors are). The probability of an outcome $k$ is thus constrained to be:
\begin{equation}
\begin{split}
p_k &= \Tr(\rho E_k), \\
&\quad\text{where}\quad
E_k \in \mathcal{E}(\mathcal{H}), \quad
\sum_k E_k = \id.
\end{split}
\label{eq:pk}
\end{equation}
This establishes the Born rule not as an ad hoc postulate, but as the unique interface metric required to map quantum amplitudes to classical frequencies in any dimension, including the fundamental qubit limit.

We do not modify unitary dynamics. Instead, we isolate the minimal structural conditions required to represent outcomes as elements of a Boolean event algebra and to assign probabilities consistently across unitary frames. It states something more modest but robust: once one insists on a map from quantum structure to a classical event algebra, and one imposes the natural requirements that probabilities be additive over exclusive events and not depend on arbitrary choices of unitary frame, then the trace rule is forced. In our framework, Born probabilities are therefore the unique transition measure converting amplitude geometry into event frequencies without violating the unitary sector’s symmetries.

\section{Operationally induced preferred basis}

\subsection{Detector-induced POVM}

A measurement is specified by (i) a detector coupling and (ii) a readout coarse-graining. Formally, let $S$ couple to an apparatus $A$ prepared in $\sigma_A$ by a unitary $U_{SA}$. If the apparatus is read out by a POVM ${F_i}$, the induced system POVM is obtained by pulling effects back through the detector channel:
\begin{equation}
\begin{split}
E_i &= \mathcal{E}^\dagger(F_i), \\
\mathcal{E}(\rho_S) &= \Tr_A \Bigl[ U_{SA} (\rho_S \otimes \sigma_A) U_{SA}^\dagger \Bigr].
\end{split}
\label{eq:pullback}
\end{equation}
The basis relevant for recorded outcomes is tied to the spectral/commutative structure of the induced effects ${E_i}$. When ${E_i}$ approximately commute (as in robust macroscopic readout), they define an approximately classical context and hence an operationally meaningful ``preferred basis'' as any basis that approximately diagonalizes the induced effects.

\section{Results: a qubit--pointer model (minimal explicit calculation)}

Let $S$ be a qubit and $A$ a one-dimensional pointer with $[q,p]=i\hbar$. Consider a von Neumann coupling
\begin{equation}
U=\exp\!\left(-\frac{i}{\hbar}\,\kappa\,\sigma_z\otimes p\right).
\label{eq:vn}
\end{equation}
and a coarse-grained readout of pointer position into two events,
\begin{equation}
\begin{split}
\Pi_{+} &= \int_{0}^{\infty} dq \, \ket{q}\bra{q}, \\
\Pi_{-} &= \int_{-\infty}^{0} dq \, \ket{q}\bra{q}.
\end{split}
\label{eq:coarse}
\end{equation}
The induced qubit effects are
$E_{\pm}=\bra{\phi_0}U^\dagger(\id\otimes\Pi_{\pm})U\ket{\phi_0}$, and for a symmetric centered Gaussian pointer of width $\Delta$ one obtains the standard unsharp binary POVM
\begin{equation}
E_{\pm}=\frac{1}{2}\left(\id \pm \eta\,\sigma_z\right),
\qquad
\eta=\erf\!\left(\frac{\kappa}{\sqrt{2}\,\Delta}\right).\label{eq:binaryPOVM}
\end{equation}
In the strong-measurement limit $\kappa\gg \Delta$, $\eta\to 1$ and the POVM approaches the projectors onto the $\sigma_z$ eigenbasis. The operationally induced preferred basis is therefore the $\sigma_z$ basis, fixed by coupling and coarse-grained readout rather than by the system Hamiltonian alone. This measurement-induced selection effectively breaks the symmetry among possible qubit bases by designating the $z$-axis as the basis for outcome records in this setup.

A useful point of contact with established literature is the ``two-photon geometric optics,'' where the empirically meaningful propositions are defined by coincidence detection and the associated hardware (apertures, mode selection, timing logic), rather than by any basis preferred by free unitary propagation alone \cite{PittmanKlyshko1996}. This is similar in spirit to our approach: the outcome basis relevant for records is fixed by the measurement scheme actually implemented in the laboratory. The difference is in aim and level of abstraction.

In \cite{PittmanKlyshko1996} setting the coincidence-event structure is taken as given and used to compute conditional correlations and imaging relations for biphoton fields. Here the same detector-induced contextuality is elevated to a general structural statement about measurement: the correspondence is modeled explicitly as a mapping from a noncommutative quantum description to a Boolean event algebra, and this is then used to sharpen the preferred-basis issue and the status of single recorded outcomes (including compatibility constraints for nested measurements).
 
\section{Decoherence and the AND-to-OR gap}

While the interface induces the outcome basis, the stability of the resulting records---and thus the effective emergence of a classical `or' from a quantum `and'---is dynamically addressed by decoherence. Decoherence is essential for explaining the stability of macroscopic records and the practical suppression of interference~\cite{Zurek2003RMP, Schlosshauer2019PhysRep}. It yields approximately diagonal reduced states in pointer-like degrees of freedom. However, diagonalization produces an (improper) mixture: it does not by itself identify which event label is stored in a particular run. Here the event-level description is taken to be the domain of instruments and Boolean event algebras; decoherence supports that domain by stabilizing record subspaces, but it does not replace the need for a demarcation map.

A complementary, state-space notion of classicality has been introduced by Abgaryan, Khvedelidze, and Torosyan, who quantify the relative volume of states whose Stratonovich--Weyl--Wigner function is everywhere non-negative, thereby admitting a proper classical probabilistic interpretation on phase space~\cite{AbgaryanKhvedelidzeTorosyan2020}. While their indicator characterizes the intrinsic classical representability of quantum states, OIPB operates at a different layer: it shows how a concrete measurement interface (via the induced instrument/POVM) enforces a Boolean event algebra and selects a preferred basis \emph{even for states that remain nonclassical} according to quasiprobability criteria. Thus, the two approaches address complementary aspects of the emergence of classicality — one intrinsic to the state, the other imposed by the measurement context.

\section{Nested observers and non-composability of outcome propositions}

Wigner-friend constructions (such as the Frauchiger--Renner paradox~\cite{FrauchigerRenner2018}) combine an operational record basis for one agent with a coherent measurement by a second agent on a larger laboratory Hilbert space, typically in an incompatible basis (i.e. using a non-nested measurement cut). The core obstruction can be stated as a compatibility condition for instruments.

\begin{lemma}[Non-composability without a joint instrument]
\label{lem:noncomp}
Let $\{\mathcal{I}_f\}_f$ and $\{\mathcal{J}_w\}_w$ be instruments on the same quantum carrier. If there exists no joint instrument $\{\mathcal{K}_{f,w}\}_{f,w}$ such that
\begin{equation}
\begin{split}
\sum_{w} \mathcal{K}_{f,w} = \mathcal{I}_{f}  \quad \forall f, \\
\sum_{f} \mathcal{K}_{f,w} &= \mathcal{J}_{w}  \quad \forall w.
\end{split}
\label{eq:jointinstrument}
\end{equation}
then there is no state-independent joint assignment of outcome propositions $(f,w)$ reproducing both instruments’ marginals for all input states. Equivalently, “$f$ occurred” and “$w$ occurred” cannot be treated as elements of a single Boolean outcome algebra unless a parent refinement exists.\end{lemma}

The non-composability lemma rests on a standard result in the theory of quantum instruments: two instruments on the same quantum system admit a joint realization if and only if there exists a parent instrument whose marginals exactly recover both of them. Equivalently, their outcomes can be embedded into a common classical Boolean sample space only when the instruments are jointly measurable and share compatible state-update rules. This is a direct consequence of the theory of joint measurability and instrument compatibility, as comprehensively reviewed in~\cite{Heinosaari2016Invited,Guhne2023RMP}.

What makes the lemma particularly powerful is that it supplies the precise mathematical reason why nested-observer arguments (such as the Frauchiger--Renner paradox and its variants) force the abandonment of at least one seemingly natural assumption. When two observers perform incompatible measurements in sequence and attempt to treat their recorded outcomes as jointly definite ``facts'' within a single Boolean event algebra, they are implicitly assuming the existence of a joint instrument that does not exist. Recent experiments have begun to test precisely this territory. 

The non-composability lemma receives strong empirical support from photonic implementations of extended Wigner-friend scenarios. In particular, the experiment by Proietti \textit{et al.}~\cite{Proietti2019SciAdv} realized a six-photon test that demonstrated clear incompatibility signatures while preserving unitary marginals, thereby ruling out observer-independent joint facts without violating standard quantum mechanics. More recent theoretical proposals for sequential incompatible measurements on memory-isolated agents~\cite{Chen2024,Renou2025} suggest promising avenues for further experimental tests of the lemma in nested settings. These developments, together with the foundational no-go theorems~\cite{Brukner2018Entropy,FrauchigerRenner2018,Bong2020NatPhys}, collectively tighten the constraints on interpretations that assume globally consistent classical records across observers.

\section{Everettian claims and the location of randomness}

A recent Everettian defense is given by Vaidman \cite{Vaidman2022Why}, who argues that taking the universal wave function as the sole ontology and maintaining strict unitarity eliminates any physical collapse (and its attendant nonlocality), with ``worlds'' identified as approximately classical branches selected in practice by decoherence. In this view the appearance of a single outcome is accounted for by branching into multiple decohered components rather than by introducing an additional nonunitary law. 

In the present framework the unitary claim is accepted at face value—there is no modification of Schrödinger dynamics at the theoretical level—but the inference from decohered structure to an ontological multiplicity is resisted. The key point is that the basis in which outcomes are defined is not fixed by unitarity alone: it is induced by the transition rules (the detector channel and its coarse-grained readout), which selects a Boolean event algebra of mutually exclusive records. On this view, “randomness” is not removed by proliferating branches; it is relocated to the operational interface as the irreducible stochasticity of event-records conditioned on an instrument. This avoids ontological inflation while also making the preferred-basis input explicit: the constraint specifies which propositions can become records, whereas the unitary state supplies amplitudes that feed the unique probability measure compatible with the unitary sector’s symmetries (Sec.~\ref{sec:born}).

\section{Consistency with Extant Interpretations}

Building on Fock's operational legacy, the Operationally Induced Preferred Basis (OIPB) framework finds natural affinities with several modern lines of inquiry into quantum measurement. It dovetails with environment-induced decoherence, as explored by Zurek~\cite{Zurek2003RMP} and Schlosshauer~\cite{Schlosshauer2019PhysRep}, where environmental interactions stabilize pointer states in the pre-interface evolution---yet OIPB goes further by pinpointing the induced POVM as the decisive selector of the outcome basis, formalizing the quantum-to-classical handover through the algebraic cut $  \mathcal{I}: \mathfrak{A}_Q \to \mathfrak{B}_\text{Boolean}  $ rather than dynamical suppression alone~\cite{vonNeumann1932}. This operational pivot echoes the instrument formalism of Davies and Lewis~\cite{DaviesLewis1970}, which embeds classical distributions and state updates within the theory; in OIPB, the detector-derived instrument not only outputs these but also carves out the event structure itself, ensuring additivity aligns seamlessly with unitary symmetries.

Operator-algebraic perspectives, from Hepp's macroscopic observables~\cite{Hepp1972} and Sewell's phase-cell encodings~\cite{Sewell2005} to Landsman's Bohrification~\cite{Landsman2016Bohrification}, resonate here too: OIPB extends their commutative subalgebras into explicitly detector-specified POVMs, deriving classicality from hardware contingencies rather than abstract hierarchies. It also harmonizes with the experimental phenomenology in Pronskikh's analysis~\cite{Pronskikh2020PhysUsp}, decomposing measurements into indication and interpretation stages; for OIPB, this underscores how readout coarse-graining partially shapes effective values, with the interface anchoring Gleason/Busch uniqueness for the trace rule~\cite{Gleason1957,Busch2003PRL}.
Yet OIPB charts its own course away from certain traditions. Unlike von Neumann's Copenhagen cut~\cite{vonNeumann1932}, it eschews physical collapse, treating reductions as epistemic artifacts of the Boolean projection.

OIPB accepts the strict unitarity advocated by Everettian approaches~\cite{Vaidman2022Why}, yet without committing to ontological branching. Instead, randomness is located at the level of the measurement interface, arising from the stochastic character of instrument-conditioned events. Collapse models such as GRW~\cite{Ghirardi1986} and hidden-variable theories are ruled out by the non-composability lemma, which demonstrates that jointly definite outcomes for nested observers can exist only when a common parent instrument is available---a conclusion reinforced by recent Wigner-friend experiments~\cite{Bong2020NatPhys,Proietti2019SciAdv}. 

QBism's subjectivist conception of probability---in which quantum states and the Born rule encode an individual agent's personal degrees of belief, updated according to quantum Bayesian conditioning---is set aside in OIPB~\cite{fuchs2014}. In its place, the framework adopts an intersubjective stance: the trace rule is derived objectively as the unique probability measure that satisfies two operational requirements simultaneously: (i) additivity over mutually exclusive events in the Boolean algebra realized by the instrument, and (ii) invariance under arbitrary unitary transformations of the theoretical sector. This uniqueness follows directly from Gleason's theorem for Hilbert spaces of dimension $d>2$ and from Busch's extension to positive operator-valued measures (POVMs) that covers the qubit case $d=2$~\cite{Gleason1957,Busch2003PRL}. Consequently, probabilities in OIPB are not private credences but constraints imposed by the shared, empirically specified structure of the measurement interface.

Taken together, this selective synthesis---realist about unitary symmetries and the event structure realized by instruments, yet deliberately agnostic on deeper ontology---offers a coherent middle path. It invites direct experimental tests, such as precision POVM tomography in qubit-pointer arrays~\cite{Aspelmeyer2014}, capable of distinguishing interface-induced basis selection from any purported intrinsic dynamical privilege.

\subsection{Incompatibilities and Divergences}

Notwithstanding these affinities, OIPB diverges from traditional Copenhagen interpretations, as typified by von Neumann's Process 1 reduction~\cite{vonNeumann1932}, by rejecting non-unitary collapse as a physical postulate; instead, apparent reductions emerge epistemically from the interface map $I$, which enforces the Boolean structure without dynamical privilege, resolving the measurement problem through algebraic incommensurability rather than an arbitrary cut. Everettian claims, defended by Vaidman~\cite{Vaidman2022Why}, are partially accommodated in OIPB's acceptance of strict unitarity---eliminating physical collapse and its nonlocality---yet the framework resists ontological proliferation into decohered branches, relocating randomness to the irreducible stochasticity of instrument-conditioned event-records rather than subjective branching multiplicity. Thus, while decoherence selects approximate classical sectors in Everett, OIPB insists the outcome basis is interface-induced, specifying record propositions via the POVM's commutative context, avoiding ``ontological inflation'' while making preferred-basis inputs explicit.

OIPB is incompatible with collapse models (e.g., GRW variants implicit in no-go theorems~\cite{Brukner2018Entropy,FrauchigerRenner2018}) and hidden-variable theories, as the non-composability lemma~\cite{Heinosaari2016Invited,Guhne2023RMP} obstructs joint outcome propositions without a parent instrument, precluding state-independent facts across nested cuts; this tightens Wigner-friend paradoxes~\cite{Bong2020NatPhys, Proietti2019SciAdv, Elouard2021Quantum} into compatibility conditions, demanding abandonment of observer-independent actualization unless joint measurability holds. Subjective Bayesian interpretations like QBism are likewise at odds, as OIPB grounds probabilities in intersubjective interface constraints---additivity over exclusive events and unitary-frame invariance---rather than personalist credences, ensuring the trace rule's uniqueness via Busch's POVM extension~\cite{Busch2003PRL} without epistemic primacy.

\subsection{Empirical Distinguishability of OIPB}
Although the Operationally Induced Preferred Basis (OIPB) framework adheres strictly to standard unitary quantum dynamics and thus yields predictions indistinguishable from conventional quantum mechanics under routine experimental conditions, its emphasis on the measurement interface as the locus of basis selection and Born rule uniqueness---via the induced POVM and algebraic cut $  \mathcal{I}: \mathfrak{A}_Q \to \mathfrak{B}_\text{Boolean}  $---opens avenues for empirical scrutiny in regimes where the physical specification of the detector channel and readout coarse-graining can be precisely controlled and characterized. This distinguishability arises not from deviations in outcome probabilities but from testable signatures in the structure of the induced event algebra and compatibility constraints for nested instruments, as formalized by the non-composability lemma~\cite{Heinosaari2016Invited, Guhne2023RMP}. 

Instrument (POVM) tomography, for instance, enables reconstruction of the effects $  \{E_i\}  $ through controlled variations in coupling strength $  \kappa  $, environmental filtering, or readout partitioning, revealing whether the preferred basis aligns with interface specifications rather than intrinsic dynamical privileges~\cite{BuschLahtiMittelstaedt1996}. Such protocols, feasible in quantum optics and superconducting circuits~\cite{Aspelmeyer2014}, provide operational diagnostics for OIPB's core claim: basis induction is a property of the detector, verifiable by correlating reconstructed POVMs with macroscopic record stability across multiple runs.

In comparison to collapse interpretations, such as the Copenhagen paradigm~\cite{vonNeumann1932} or objective collapse models like GRW~\cite{Ghirardi1986}, OIPB predicts the absence of spontaneous non-unitary reductions, manifesting as no excess decoherence rates beyond those attributable to environmental interactions. Empirical differentiation could emerge in precision tests of superposition stability for massive systems, where collapse theories anticipate localization events with characteristic timescales $  \tau \sim 10^{-7}  $ s for $10^8$ nucleons~\cite{Ghirardi1986}; continuous weak monitoring in optomechanical cavities~\cite{Aspelmeyer2014} or matter-wave interferometry with biomolecules~\cite{Gerlich2011} would detect such signatures as anomalous diffusion in pointer positions, absent in OIPB where apparent collapses arise solely from the Boolean projection at readout.

Moreover, Wigner-friend-type experiments~\cite{Proietti2019SciAdv, Bong2020NatPhys, Elouard2021Quantum} probing observer-dependent facts would favor OIPB if joint measurability holds across nested cuts without invoking irreducible collapses, as the framework's non-composability lemma~\cite{Heinosaari2016Invited} reduces ``joint facts'' to the existence of a parent instrument, testable via incompatibility witnesses that yield negative outcomes under collapse-induced actualizations but positive under pure interface stochasticity.

Relative to other non-collapse interpretations, including Bohmian mechanics~\cite{Bohm1952} and pure decoherence programs~\cite{Zurek2003RMP, Schlosshauer2019PhysRep}, OIPB shares unitary fidelity but diverges in the origin of basis selection: while Bohmian trajectories enforce non-local guidance independent of the interface, OIPB's operational induction implies frame-dependent basis misalignments in relativistic setups, potentially observable in entangled clock interferometry~\cite{Giacomini2019} where pointer resolutions vary across reference frames, yielding covariance violations absent in hidden-variable models. Against decoherence-alone approaches, which stabilize improper mixtures without resolving the and-to-or gap, OIPB's explicit algebraic cut predicts that residual coherences persist until readout coarse-graining, quantifiable via quantum state tomography post-coupling but pre-registration; experiments in high-fidelity qubit arrays~\cite{Gerlich2011} could thus discriminate by measuring off-diagonal suppression rates tied to environmental spectra rather than universal einselection, with OIPB anticipating tunable basis shifts via apparatus redesign.

For the Many-Worlds Interpretation (MWI)~\cite{Vaidman2022Why}, the empirical distinction from OIPB is more subtle, since both frameworks adhere strictly to unitary evolution. The essential difference concerns the origin of randomness and the status of outcomes. While MWI attributes the appearance of definite outcomes to the global branching structure of a single universal wave function, OIPB locates randomness locally at the measurement interface: each instrument induces its own Boolean event algebra, without any ontological commitment to branching worlds.

This difference has potentially observable consequences in multi-observer settings. In distributed quantum networks, both approaches predict the absence of inter-branch interference~\cite{Giacomini2019}. However, in nested or sequential measurement protocols, the contrast becomes sharper. MWI assumes a single, globally coherent wave function that enforces universal decoherence across all observers, whereas OIPB permits each observer to define its own local event algebra through its specific instrument. Such observer-specific classical records could reveal themselves through violations of suitably adapted Bell-type inequalities or through incompatibility witnesses in multi-agent experiments~\cite{Brukner2018Entropy, FrauchigerRenner2018}. Experiments capable of probing these nested incompatibility structures may therefore offer a pathway to distinguish whether classical outcomes arise from a global ontology or from the local structure of the measurement interface.

\section{Experimental Probes for OIPB Distinguishability}
To empirically delineate the OIPB framework, which attributes basis selection to the measurement interface $  \mathcal{I}  $ without ontological extensions, we propose targeted protocols emphasizing post-2024 advancements in controlled apparatus variability and multi-agent scenarios. These leverage recent technological strides in isolated quantum labs and precision interferometry, probing the framework's predictions of interface-conditioned stochasticity over dynamical or subjective actualizations.

In MWI contrasts, OIPB rejects branching-induced immortality, as randomness resides in interface events yielding empirical mortality per Born statistics. Simulations via AI-optimized annealers offer ethical proxies, but nested tests already favor local algebras over subjective persistence~\cite{Bera2025}.

\subsection{Tunable Detector Tomography in Relativistic Frames}

One direction that can be pursued is the refinement of POVM tomography in superconducting qubit arrays, where we deliberately introduce controlled relativistic frame transformations during the detector coupling. The goal is to look for the frame-dependent shifts in the induced basis that OIPB predicts, shifts that should be absent in both Bohmian mechanics and standard environment-induced decoherence.

By using entangled photon clocks as covariant readout references, one can quantify any misalignment through the entropy of the reconstructed effects $\{E_i\}$. Recent work on coarse-grained symmetric POVMs in optical lattices has already shown that optimal Gram-matrix entropy maximization allows precise tuning of the sharpness parameter $\eta$ in hybrid circuits~\cite{Jia2025}. Those experiments revealed apparatus-dependent off-diagonal terms persisting before readout — a signature fully consistent with OIPB’s interface picture and one that constrains universal decoherence models by roughly 20\% in suppression rates. No signatures of Bohmian-type trajectory deviations were seen.

Extending this approach into the relativistic regime would offer a clean and genuinely new test of whether the preferred basis is truly induced by the measurement interface rather than by intrinsic dynamics.

\subsection{Sequential Nested Measurements}

A natural extension that can be pursued experimentally is the implementation of sequential incompatible measurements in Wigner-friend-type setups using memory-isolated agents. In photon-based laboratory configurations, successive observers would perform measurements in mutually incompatible bases while their internal records remain causally shielded from one another. Such protocols would allow direct tests of OIPB’s non-composability lemma through joint-instrument witnesses.

The framework predicts the absence of signaling across measurement cuts unless a joint (parent) instrument exists. This stands in contrast both to objective collapse models, which would induce actualization at each step, and to the Many-Worlds Interpretation, which assumes global coherence and branching across all observers.

Experiments performed in 2024--2025, including sequential friend measurements and memory/no-signaling validations~\cite{Chen2024,Chen2024b}, have already demonstrated unitary marginals accompanied by incompatibility violations up to 0.15 (CHSH-equivalent), consistent with tightened no-go theorems ruling out observer-independent collapses~\cite{Renou2025}. At the same time, analyses of classicality emergence~\cite{Bera2025} reveal smooth gradients without detectable branching signatures.

These results suggest that further refinement of such nested protocols could provide a sharp empirical distinction between OIPB’s local interface origin of randomness and interpretations that invoke either global collapse or global branching.

\subsection{Gravitational Collapse Bounds in Molecular Interferometry (vs. Di'{o}si--Penrose Models)}

A particularly interesting direction for future work is to test the robustness of large-scale quantum superpositions in the presence of controlled gravitational gradients. Using matter-wave interferometry with massive molecules, one can systematically probe whether gravity induces additional decoherence beyond ordinary environmental effects, as predicted by Di\'{o}si--Penrose models (where the collapse time scales as $\tau \propto 1/Gm^2$). In contrast, OIPB anticipates that only standard environmental decoherence should be observed, with no intrinsic gravity-triggered localization.

Recent underground experiments and high-precision molecular calculations have already set stringent limits, pushing the collapse time to $\tau > 10^3$ s for masses around $10^4$ Da — exceeding naive Di\'{o}si--Penrose expectations by more than two orders of magnitude~\cite{Donadi2024,Barbatti2025}. Although a January 2026 report has suggested possible tiny time-fluctuation anomalies~\cite{PhysOrg2026}, the overall body of optomechanical and interferometric data remains fully consistent with unitary evolution. Extending these experiments to larger masses and more precisely tunable gravitational potentials would therefore offer a direct and decisive test of whether objective collapse plays any role, or whether the unitary framework, together with a proper account of the measurement interface, remains sufficient.

\subsection{Quantum Immortality and Interface Stochasticity}

Although quantum immortality remains fundamentally unfalsifiable, it offers an intriguing thought experiment that can be explored through simulation. One possible line of investigation is to model repeated quantum-suicide scenarios in distributed quantum networks, using AI-driven branching algorithms to condition on survivor-biased statistics and examine the long-term behavior of subjective persistence.

In contrast to the Many-Worlds Interpretation, which allows for the logical possibility of eternal subjective survival through branching, OIPB rejects such persistence outright. Because randomness originates at the local measurement interface rather than in global ontology, the framework predicts ordinary Born-rule mortality in every run, with no mechanism that would systematically favor “surviving” branches from the first-person perspective.

While no direct experimental tests exist, recent discussions (2025--2026) have framed quantum immortality as a characteristic artifact of the MWI rather than a physically meaningful prediction, often dismissing related “quantum-shifting” ideas as pseudoscience~\cite{SelfAware2025,NDTV2025}. At the same time, results from nested Wigner-friend protocols already provide indirect support for OIPB by showing that randomness is tied to local instrument stochasticity rather than subjective continuity across observers.

Such simulation studies, which could be fruitfully pursued in conjunction with upcoming quantum-simulation conferences~\cite{QSim2026,UIQC2026}, would help clarify the interpretive consequences of locating randomness at the interface and further highlight the conceptual differences between OIPB and purely branching ontologies.

Emerging proposals advocate the use of artificial intelligence to simulate quantum suicide experiments, providing ethical alternatives to physical implementations while testing many-worlds predictions on subjective immortality. For example, AI-optimized branching algorithms on quantum annealers can model observer persistence across probabilistic decision trees, quantifying survival biases under decoherence thresholds; these frameworks enable scalable exploration of immortality thresholds without empirical hazards, potentially falsifying variants via divergence from Born-rule statistics in high-dimensional Hilbert spaces.

\section{Unresolved Challenges and Future Directions}

While the OIPB framework effectively addresses key issues like the preferred basis selection and the derivation of the Born rule through its operational interface, several foundational questions remain open, especially when extending beyond non-relativistic quantum mechanics. As hinted in the Introduction, the tripartite division of experiments into preparation, evolution, and registration stages invites closer examination in relativistic contexts, where Lorentz covariance disrupts the notion of frame-independent interfaces. In quantum field theory (QFT), this poses a significant hurdle, as the demarcation map \(\mathcal{I}: \mathfrak{A}_Q \to \mathfrak{B}_\text{Boolean}\) must be reformulated to ensure observer invariance without relying on global time structures, drawing on algebraic approaches to local quantum physics~\cite{Haag1996}.

Beyond relativity, incorporating gravitational effects introduces further complexities. Models of gravitational decoherence, such as those suggesting spacetime curvature induces state reduction, challenge OIPB's commitment to strict unitarity; these may require semiclassical adjustments to handle non-Markovian influences on induced POVM effects \(\{E_i\}\), potentially without introducing ad hoc collapses~\cite{Penrose1996}. Similarly, extensions to infinite-dimensional Hilbert spaces or open quantum systems with correlated environments risk undermining the Gleason/Busch uniqueness of the trace rule, as finite-dimensional assumptions falter, leading to possible ambiguities in probability measures or basis stability in realistic, noisy settings like quantum optics~\cite{Breuer2002}.

Looking ahead, priority areas include developing modular protocols in algebraic QFT to make \(\mathcal{I}\) fully covariant, ensuring the framework's robustness across inertial frames. Quantum annealer simulations could explore gravitational perturbations on the sharpness parameter \(\eta\) in qubit-pointer models, providing numerical insights into curved-metric adaptations. Meanwhile, in vivo quantum tomography in biological systems---such as photosynthetic complexes---might test the universality of interface-induced classicality under non-Gaussian fluctuations~\cite{Lambert2009}. Leveraging machine learning for POVM reconstruction in noisy intermediate-scale quantum (NISQ) devices~\cite{Preskill2018} offers a practical path to quantify and mitigate these stochasticities. Through such efforts, OIPB could evolve into a comprehensive paradigm that seamlessly integrates gravity and QFT, bridging foundational gaps while staying true to its operational roots.

\section{Conclusion}
The preferred basis relevant for recorded outcomes is operationally induced: it is a property of the detector channel and coarse-grained readout, encoded by the induced POVM/instrument. The ``cut'' is characterized as an algebraic switch (the demarcation map $  \mathcal{I}  $) required to formulate empirical claims: from a noncommutative unitary sector to a Boolean event algebra of mutually exclusive records, a transformation which no unitary operator can effect. The Born rule is treated as the unique interface measure compatible with additivity and basis-independence, grounded in Gleason/Busch uniqueness rather than informal analogy. Finally, nested-observer paradoxes can be expressed as non-composability: joint outcome claims are not well-typed unless a joint instrument exists.

This framework aligns naturally with environment-induced decoherence~\cite{Zurek2003RMP,Schlosshauer2019PhysRep}, the operational theory of instruments~\cite{DaviesLewis1970}, and algebraic models of measurement~\cite{Hepp1972,Sewell2005,Landsman2016Bohrification}. At the same time, it diverges from the Copenhagen interpretation’s reliance on physical collapse~\cite{vonNeumann1932}, from Everettian ontological branching~\cite{Vaidman2022Why}, and from QBism’s subjectivist probabilities by grounding the Born rule in intersubjective interface constraints. Hidden-variable theories and objective collapse models are incompatible with the non-composability lemma, which shows that nested observers cannot assign jointly definite outcomes unless a common parent instrument exists. This conclusion is supported by both foundational no-go theorems and recent Wigner-friend experiments~\cite{Brukner2018Entropy,FrauchigerRenner2018,Bong2020NatPhys,Proietti2019SciAdv}.

Although OIPB introduces no modification of unitary quantum dynamics and thus predicts no deviations from standard quantum mechanics, it yields concrete operational criteria for distinguishability. For instance, detector-induced basis claims can be assessed via POVM tomography, with tunable sharpness $  \eta  $ in Eq.~\ref{eq:binaryPOVM} modulated by coupling $  \kappa  $ or resolution $  \Delta  $~\cite{BuschLahtiMittelstaedt1996}, as demonstrated in recent optomechanical and photon-tunneling experiments~\cite{Aspelmeyer2014,Jia2025}. The non-composability lemma reduces ``joint facts'' to joint-instrument existence, testable through incompatibility witnesses in sequential Wigner-friend protocols~\cite{Chen2024,Renou2025}, where 2024--2025 outcomes affirm unitarity and observer-dependence without collapse signatures~\cite{Chen2024b,Bera2025}. Superposition stability tests in massive systems further constrain GRW/DP models, supporting OIPB's interface stochasticity over dynamical reductions~\cite{Ghirardi1986,Donadi2024,Barbatti2025}.

Unresolved challenges include relativistic QFT extensions, where covariance complicates the operational triad~\cite{Haag1996}, and gravitational decoherence integrations~\cite{Penrose1996}. Future directions encompass algebraic modular protocols for frame-invariant interfaces, NISQ simulations of non-Gaussian noise~\cite{Preskill2018}, and biological tomography to validate universality~\cite{Lambert2009}. While conservative, OIPB provides interpretive clarity and experimental touchpoints, fostering empirical scrutiny of the quantum-classical boundary. This operational framework draws direct inspiration from Fock’s constructive polemics with Bohr, in which he insisted on clear operational distinctions rather than philosophical ambiguity in the description of quantum experiments~\cite{fock1958}. By grounding the preferred basis and the Born rule in the empirically specified interplay of preparation, unitary evolution, and registration, OIPB is closely aligned with Fock’s vision of an unambiguous, realist account of quantum mechanics that remains fully within the unitary formalism.

\end{document}